# Solar Energy Conversion and the Shockley-Queisser Model,
*a Guide for the Perplexed*

Jean-Francois Guillemoles[1], Thomas Kirchartz[2,3], David Cahen[4], and Uwe Rau[2]

[1]CNRS, UMR 9006, Institut Photovoltaique d'Ile de France (IPVF), Palaiseau, France

[2]IEK5-Photovoltaik, Forschungszentrum Jülich, 52425 Jülich, Germany

[3]Fac. of Engineering and CENIDE, Univ. of Duisburg-Essen, Carl-Benz-Str. 199, 47057 Duisburg, Germany

[4]Department of Materials and Interfaces, Weizmann Institute of Science, Rehovoth 76100, Israel

**e-mails for correspondence:**
Jean-Francois.GUILLEMOLES@cnrs.fr ; david.cahen@weizmann.ac.il ; t.kirchartz@fz-juelich.de; u.rau@fz-juelich.de

**Abstract**   The Shockley-Queisser model is a landmark in photovoltaic device analysis by defining an ideal situation as reference for actual solar cells. However, the model and its implications are easily misunderstood. Thus, we present a guide to help understand and avoid misinterpreting it. Focusing on the five assumptions, underlying the model, we define figures of merit to quantify how close real solar cells approach each of these assumptions.

## Introduction

In 1961 Shockley and Queisser[1] (SQ) analyzed the limits of photovoltaic energy conversion, using the basic thermodynamic principle of detailed balance instead of phenomenological approaches, used earlier.[2-4] The final result of their analysis is commonly referred to as the 'SQ-limit'. While arguably the most important theoretical contribution to photovoltaic energy conversion, the paper also relies on a highly idealized model for solar cells, using substantially simplifying assumptions. Therefore, only within the assumptions of their model (denoted the SQ-model in the following) does the term 'SQ-limit' make sense. In view of the emergence of promising new photovoltaic absorber materials and devices with very high efficiencies[5] with various claims of 'exceeding or approaching the SQ-limit',[6,7] we will critically discuss the connection of the SQ-model to real world solar cells and will explain what 'close' to the SQ-model means.

First, we briefly describe the SQ-model in its initial form by illustrating its three fundamental steps, noting the energy losses associated with each of these. We then describe the five assumptions that are the essence of the model (Table 1). Subsequently, we examine how each of these assumptions compares to more realistic situations, discuss experimentally measurable figures of merit (FoMs) and quantify how real-world solar cells differ from the ideal model, thereby providing guidelines for effective use of the SQ-model.

*Description of SQ-model and its assumptions*

Figure 1a shows the setting of the SQ-model in its original form. The solar cell interacts with the surroundings by exchanging light particles (photons) with the sun and with the ambient. Furthermore, the cell exchanges electrons with the external electrical circuit and heat with a temperature reservoir to maintain the cell temperature $T_{cell}$ constant and equal to the ambient temperature $T_{amb}$, such that without solar irradiation cell and ambient are in thermodynamic equilibrium. We are *not* considering light concentration or a restriction of the angle of optical interaction of the cell with the ambient. With this setting we aim at a simple picture that, however, is compatible with more elaborate, thermodynamic descriptions of the solar cell working principles.[8-10]

The photovoltaic absorber in the SQ-model is a semiconductor, described by two groups of electronic energy levels that extend throughout the material (cf. Fig. 1b). The lower one–called *valence band*, VB – is filled with electrons, while the upper one – the *conduction band*, CB – is initially empty. The energy difference between the edges $E_C$ and $E_V$ of the conduction and valence bands is the band gap energy $E_g$ (Fig. 1b). If sufficient energy is supplied, an electron can be promoted from VB to CB, leaving behind an empty state in the VB. The electron in the CB and the empty state in the VB (called hole) now behave as "free" charge carriers that can move in their respective bands (cf. Fig 1b).

The solar radiation is described by photons with a distribution of energies, the solar spectrum (NB: while differing situations may require different solar spectra, the present analysis remains valid for any). Depending on their energy, these photons may have enough energy to create free electrons and holes. These electron-hole pairs, generated by photon absorption, can also annihilate themselves by releasing their energy in the form of photons (*radiative recombination*).

The interaction of the photon with the semiconductor and the photovoltaic action in the solar cell proceed in three stages (1-3, as sketched in Fig. 1b), with relevant time-scales shown in Fig. 1c and Tab. 1.

(A – optical) absorption of a photon with creation of an electron-hole pair,

(B – thermal) relaxation of this electron-hole pair towards $E_C$ and $E_V$,

(C – electronic) extraction of the electron and hole at two different contacts, or their radiative recombination followed by emission of a photon.

Note that the extraction of charge carriers at different contacts requires a built-in asymmetry that separates electrons from holes and that may be achieved electrostatically via a pn-junction. However, while the pn-junction is featured in the title of the original SQ paper and used for illustration purposes in Fig. 1b, it is by no means a necessary requirement for an efficient solar cell.[8,11]

Within the SQ-model, the three stages are defined by 5 assumptions:

*OPTICAL:* It is assumed that for impinging photons with photon energy $E > E_g$, the photon is absorbed, whereas for $E < E_g$ photons do not interact with the solar cell at all (**assumption 1**), i.e., the absorptivity $A(E)$, a measure of the material's ability to absorb radiation, is a step function, 0 for $E < E_g$ and 1 for $E > E_g$. Hence, the first energy loss is due to the solar cell's transparency for $E < E_g$. Also, absorption of a photon with $E > E_g$ generates precisely one electron-hole pair that contributes to the short-circuit current $J_{SC}$ (**assumption 2**).

*THERMAL:* The electron-hole pair loses (to the absorber material's lattice) all excess energy above $E_g$, i.e., the pair relaxes in sub ps timescales to the average energy of a thermalized electron-hole pair (in thermal equilibrium with the cell at $T_{cell}$). The underlying **assumption 3** is that in the solar cell all relaxed electron-hole pairs are at the same temperature $T_{cell}$.

*ELECTRONIC:* At this point, one of two things can happen to the electron-hole pair: Either the electron and the hole are collected at their respective contacts or they recombine radiatively by emission of a photon (this is the only allowed recombination mechanism, **assumption 4**). Note that for an actual recombination, the photon must be emitted to the ambient because reabsorption of the photon in the solar cell creates a new electron-hole pair and restarts the whole process from the beginning (photon recycling).[12] Both phenomena (recombination and extraction) occur typically on ns to µs timescales – collection usually being faster than recombination. Emission of photons from radiative recombination causes a photon flux, described by the product of the absorptivity and the black body spectrum of the solar cell. The connection between absorption and emission of a semiconductor is a result of the principle of *detailed balance*, which states that every microscopic process must have the same rate as its inverse process in thermal equilibrium.[13] Otherwise, thermal equilibrium could not be reached.

The SQ-theory, assumes that the rate constants derived from the principle of detailed balance for thermal equilibrium are also valid in the non-equilibrium situation.[8]

The collection of photogenerated carriers implies that the solar cell has two different external contacts that can support an external voltage $V$ and carry a current $J$. **Assumption 5** states that each contact is ideal because each exchanges only one carrier type (electrons or holes) with the absorber (*selective contact*) and because it has negligible resistance. Nevertheless, collection of the electron-hole pair implies that its total energy is reduced from the band gap energy $E_g$ to $qV$, the electrical work of transferring a charge carrier between the contacts. We denote this loss of potential energy 'isothermal dissipation' as it generates heat in the solar cell *without* a change of the temperature of electrons and holes (unlike during thermalization) and the carriers do *not* recombine (unlike during radiative recombination). Note that this loss can be subdivided further into different reversible and non-reversible contributions.[10,14] We have listed four energy loss mechanisms in the SQ-model in Table I and illustrated in Fig. 2a which also shows the maximum efficiency of 30 % (referring to a 5800 K black body solar spectrum, likewise a value of 33 % would hold for the more complex terrestrial spectrum) at an optimum band gap energy. Fig. 2b shows the specific share of the four energy losses for a specific band gap energy. It also illustrates how the output power is maximized by the proper choice of the voltage minimizing the sum of emission and isothermal dissipation losses.

The efficiencies depicted in Fig. 2a based on the SQ-model represent a separation line between the so-called third generation photovoltaics[15] (where higher efficiencies can be achieved by bypassing at least one of the 5 assumptions in a way to reduce power losses) and the 'normal' single junction solar cells with efficiencies below the SQ-case (for which the 5 assumptions are approached but not reached resulting in higher power losses than implied by the SQ model, as described by the FoMs, below).

**Accounting for losses in real, single absorber solar cells**

We now describe departures from the ideal SQ-model for real-world single junction solar cells by successively relaxing the 5 assumptions. Given that quantitatively accounting for the losses in a physically meaningful way becomes quite detailed and technical in parts we encourage the casual reader to jump directly to the "Consequences" section. For readers preferring to quantitatively understand the consequences of these departures from the SQ-model, we express the current versus voltage (*JV*) characteristics of the solar cell in terms of a simple balance equation (Eq. (1) in the Box) expressing that the net total electrical current $J$ extracted from the cell equals the short circuit photocurrent $J_{SC}$, (a gain) minus the recombination or diode current (a loss). The diode current is a product of a prefactor (the saturation current $J_0$ that measures the recombination loss) and a rectifying term varying exponentially with voltage (see Fig. 2b and c). From equation 1 all further information like the open-circuit voltage $V_{OC}$, the maximum output power $P_{max}$, and the efficiency $\eta$ are derived by simple mathematics; the Box gives the key equations. The rectifying character of the *JV*-characteristic, essential for an efficient photovoltaic device, is measured by the fill factor *FF*, the ratio of the electrical power $P_{max}$ at the maximum power point divided by the product $J_{SC}V_{OC}$. As long as equation 1 is valid, the fill factor - at a given temperature $T_{cell}$ – is a well-defined, increasing, function of $V_{OC}$, i.e. $FF = FF_0(V_{OC})$ (equation 6, Box).[16]

The SQ-model with assumptions 1-5 yields a short circuit current $J_{SC}^{SQ}(E_g)$ and a saturation current $J_0^{SQ}(E_g)$ that only depend on $E_g$ (as material property). These two values determine the output power $P_{max}^{SQ}$, for a given $T_{cell}$ and a given solar spectrum, solely by $E_g$, as shown in Fig. 2a). As discussed in the following, relaxing assumptions 1-5 leads to power losses by changes in one or more of the three parameters $J_{SC}, J_0$ , $T_{cell}$ and also to changes of the overall *J-V* shape as discussed in the following.

*(A) OPTICAL:* No real solar cell can fulfill assumption 1, i.e, have a step function absorptivity *A(E)*, e.g. because of broadened CB and VB edges, called band tails[17] resulting from (static) structural disorder,[18] due to the existence of charge transfer states in organic semiconductors,[19] or simply because of a finite cell thickness and finite absorption coefficients. An obvious problem in this context is to define the band gap to use in the SQ model to determine $J_{SC}^{SQ}$ and $J_0^{SQ}$ that may serve as reference values. Especially for semiconductors with (static) disorder the exact definition or method to determine the band gap is problematic. To be consistent with the SQ-model we recently proposed to use the derivative of the external photovoltaic quantum efficiency with respect to photon energy as measure of photovoltaic band gap.[20] In violation of assumption 2, parasitic absorption of photons in contact layers or by free carriers (electrons) in the optical absorber reduces the average number of photogenerated electron-hole pairs per absorbed photon to < 1.

The consequence of violating assumptions 1 and 2 of the SQ-model are best understood in terms of the external photovoltaic quantum efficiency $Q_e^{PV}(E)$, i.e., the probability that a photon of energy *E*, impinging on the cell, generates an electron-hole pair that, under short-circuit conditions, is extracted at the contacts. In general we have $Q_e^{PV}(E) < A(E)$ and it is easily understood that also this leads to reduction of the short circuit current.

Thus, violating assumptions 1 and 2 decreases the short circuit current from the ideal value $J_{SC}^{SQ}$ to the real value $J_{SC}^{QE}$, determined by $Q_e^{PV}(E)$. Because of the reciprocity between $Q_e^{PV}(E)$ and the electroluminescence of a solar cell,[21] $J_0^{SQ}$ also changes to $J_0^{QE}$ (which still describes radiative loss to the ambient). It should be noted that the reciprocity applies to all optical absorbers, nanophotonic or not, so that this analysis is generally valid. This point is discussed in detail in ref. [22].

Summarizing the loss resulting from assumptions 1 and 2, we use the ratios of short-circuit currents $J_{SC}^{QE}/J_{SC}^{SQ} (= F_{SC}) \leq 1$ and radiative emission loss currents $J_0^{SQ}/J_0^{QE} (= F_{em}) \leq 1$ as two FoMs that describe departures from the ideal SQ-case. The best experimental values of $F_{SC}$ are $> 90\%$ for top laboratory cells of all commercial types (95% for best c-Si cells),[5] while typical $F_{em}$ values range from $\approx 0.1 - 0.5$ in GaAs or c-Si cells to $\ll 10^{-3}$ in amorphous or organic cells.[20] Note the different weight of $F_{em}$ entering only logarithmically in the open circuit voltage and in the efficiency equation (Eqs. 3, 5; Box) as compared to $F_{SC}$, which has an approximately linear effect on the solar cell efficiency. Thus, a 10 % loss in $F_{SC}$ implies a little bit more than 10 % loss in efficiency whereas a 90 % loss in $F_{em}$ involves a loss in $V_{OC}$ of $kT_{cell}/q \times \ln(10) \approx 60$ mV, i.e., $< 10 \%$ loss, if $V_{OC} > 600$ mV.

*(B) THERMAL:* The most likely consequence of violating assumption 3 (electron-hole pairs, absorber and contacts at the same temperature) is that the operation temperature $T_{cell}$ of real solar modules outdoors[23] is $> 300$ K (as assumed in the SQ-model) or than the standard-testing-conditions (STC) temperature $T_{cell}^{STC} = 25$ °C, used to rate efficiency. Because any recombination current (radiative and non-radiative, expressed by the respective $J_0$ -values) is thermally activated, the corresponding recombination loss increases exponentially with increasing temperature. Thus, the major effect of elevated temperature is reducing the open-circuit voltage (see Eq. (3), Box), with second order effects on other parameters. The reduction of the annual energy yield of modules outdoors is 2 – 10 % depending on technology and location.[23]

*(C) ELECTRICAL:* An equivalent circuit model for the solar cell (Fig. 1d) helps to understand the electrical losses. As long as assumptions 4 and 5 are valid, the circuit consists of a diode where the saturation current $J_0^{QE}$ is given by radiative emission. This implies that we are still in the "radiative recombination only" situation, while in practical solar cells non-radiative recombination of electron-hole pairs in the bulk and at interfaces (violation of *assumption* 4) occurs and represents a major additional loss channel. With the notable exception of the GaAs record cells, recombination losses in solar cells are entirely dominated by non-radiative recombination, Thus, violation of assumption 4 requires replacing $J_0^{QE}$ by the saturation current $J_0^{real}$, the sum of non-radiative recombination and the still unavoidable radiative one. This change generally implies an increase of recombination losses by orders of magnitude and a corresponding decrease of the open-circuit voltage (Eq. 3, Box). A common FoM, the external luminescence quantum efficiency $Q_e^{lum} = J_0^{QE}/J_0^{real}$ has been extensively analyzed for various solar cell types with peak values of $Q_e^{lum}$ = 20 % for the record GaAs devices,[24] and ~ 5% for lead-halide perovskite cells *under operating conditions*.[25] But many solar cells, including the ones based on crystalline Si, have $Q_e^{lum} \leq 1$ % implying >120 mV $V_{OC}$ losses with respect to the radiative limit. We note that in the present analysis we have assumed that the so-called diode ideality factor $n_{id}$ for non-radiative recombination is unity (as for radiative recombination).

Violations of assumption 5, stating that electrons and holes can move freely everywhere in the absorber towards/from their electron (hole) contact are, in the simplest case, described by additional resistive elements in Fig. 1d, that *change the shape* of the *JV*-curve while hardly affecting $J_{SC}$ and $V_{OC}$. As illustrated in Fig. 2c, violating assumption 5 (e.g. increased series resistance) will decrease *FF* below $FF_0$ and reduce the solar cell efficiency accordingly. Thus, an especially simple figure of merit is $F_{FF}^{res} = FF_{real}/FF_0(V_{OC}^{real})$. As presented in table II, typical $F_{FF}^{res}$ values are > 97% for record Si and GaAs solar cells, and range between over 90% for polycrystalline thin film and near 85 % for polymer-based devices.[26]

**Consequences**

Returning to our initial question, we can now use the FoMs $F_{SC}$, $F_{em}$, $Q_e^{lum}$ and $F_{FF}$ to measure how close a solar cell is to the SQ-model in the dimensions associated with 4 out of the 5 assumptions of the SQ-model ($T_{cell}$ is kept at 300 K). This procedure goes beyond a simple comparison of the photovoltaic parameters $J_{SC}$, $V_{OC}$, and $FF$ with their respective SQ-reference values.[27] We stress that $F_{SC}$, $F_{em}$  $Q_e^{lum}$, and $F_{FF}^{rec}$ should be experimentally determined *on complete devices*. The SQ-situation is approached *if all four* quantities are close to unity. Optimizing devices with respect to only one of the FoMs *cannot* be associated with the SQ-limit because such optimization often goes at the expense of the others. This issue becomes even more critical if measurements are performed on incomplete devices or on photovoltaic absorbers only. For absorber materials, the external luminescence efficiency  $Q_e^{lum}$ is frequently measured by means of photoluminescence,[28] thus replacing electrical excitation as in the proper experiment on completed devices, by optical excitation. However, optimizing $Q_e^{lum}$, e.g. by applying well-passivating (but possibly current-blocking) surface layers, may guide us in the wrong direction. Instead of maximizing one FoM against all others, we need to optimize all of them simultaneously and need to understand the trade-offs that prevent to put all FoMs close to unity at the same time. Notably, a part of the FoMs can be extracted from standard illuminated current-voltage and quantum efficiency measurements that are, e.g., tabulated in Ref. [5] for a set of recent record cells from different technologies. Table II and Fig. 3 show the FoMs $F_{SC}$ and $F_{FF}^{res}$ and the values $V_{OC}^{real}/V_{OC}^{SQ}$ and  $FF_0(V_{OC}^{real})/FF_0(V_{OC}^{SQ})$ that are used in Eq. (5) (Box) to calculate the efficiency ratio  $\eta_{real}/\eta_0$. Note that the FoMs  $Q_e^{lum}$ and $F_{em}$ cannot be obtained separately without an analysis of the luminescence emission of the device.[20] However, a presentation of different loss terms based on an incomplete set of FoMs like Fig. 3 is already useful to identify strengths and weaknesses of different technologies with a precise reference to the SQ-case.

**Acknowledgements:**
JFG thanks the French program of "investment for the future" (ANR-IEED-002-0). DC thanks the Inst. PV d'Ile de France for a visiting professorship and the Ullmann family foundation (via the Weizmann Inst) for support.

...

**Table I:** The three steps (A-C) of the SQ-model, their time scales and energy losses, associated with them, and real solar cell departures from the five SQ-assumptions (Fig.1), quantified by five FoMs (**see Box for symbols**): **1. + 2.** Relaxing assumption 1 towards a non-step function-like absorptivity $A(E)$, combined with relaxing assumption 2 (not all absorbed photons lead to electron-hole pairs that are collected) has two consequences: Firstly, the short circuit current $J_{SC}^{QE}$ of a real cell becomes *smaller* than $J_{SC}^{SQ}$, the SQ-model value. Secondly, the radiative loss current $J_0^{QE}$ becomes *larger* than $J_0^{SQ}$. Accordingly, we define two FoMs $F_{em}$ and $F_{SC}$, unity in the SQ-limit, but < 1 if assumptions 1 and 2. are not fulfilled. **3**. Deviation of the operating cell's temperature, $T_{cell}^{op}$, and $T_{cell}^{SQ}$=300K assumed in the SQ-model is expressed by $F_T = T_{cell}^{SQ}/T_{cell}^{op}$. **4**. The occurrence of non-radiative recombination (violation of assumption 4), is described by $Q_e^{lum}$, the ratio between the emitted radiation, $J_0^{QE}$ and the total recombination current $J_0^{real}$. **5**. The ratio $F_{FF}^{res}$ describes fill factor losses due to non-zero series resistance, low charge-carrier mobilities and finite parallel resistances, in violation of assumption 5.

**Table 1: Stages and Assumptions of the SQ-model:**

| Stages **A-C** of SQ-model (Time scales) - *Energy losses* | Assumptions **1-5** of SQ-model | Changes of diode parameters | Figures of merit |
|---|---|---|---|
| **A. OPTICAL** (1-10 fs) - *Loss of photons that are not absorbed* | 1. At $E_g$ absorptivity of photons in absorber switches from 0 to 1 | | |
| | 2. Exactly one electron-hole pair per absorbed photon. Each pair is collected at short circuit. | 1. $J_{SC}^{SQ} \to J_{SC}^{QE}$  2. $J_0^{SQ} \to J_0^{QE}$ | 1. $F_{SC} = J_{SC}^{QE}/J_{SC}^{SQ}$  2. $F_{em} = J_0^{SQ}/J_0^{QE}$ |
| **B. THERMAL** (0.1-10 ps) - *Loss of excess kinetic energy* | 3. Heat extraction from the carrier system such that the carrier temperature equals cell and ambient temperature | 3. $T_{cell}^{SQ} \to T_{cell}^{op}$ | 3. $F_T = T_{cell}^{SQ}/T_{cell}^{op}$ |
| **C. ELECTRONIC** (0.1-1000 ns) - *Loss by emission of photons* - *Isothermal dissipation loss during carrier collection* | 4. Electron-hole recombination is only radiative (emission of radiation) | 4. $J_0^{QE} \to J_0^{real}$ | $Q_e^{lum} = J_0^{QE}/J_0^{real}$ |
| | 5. No Ohmic losses, contacts are perfectly selective | 5. $FF_0(V_{OC}^{real}) \to FF_{cell}$ | 4. $F_{FF}^{res} = FF_{cell}/FF_0(V_{OC}^{real})$ |

**Table II:** The figures of merit discussed in table I for a range of record cells listed in supplementary table 1 and 2 in ref. [5]. As a reference band gap, we use the photovoltaic band gap extracted from solar cell quantum efficiency data as described in ref. [20]. From the typically published data on solar cells, the FoMs as $F_{SC}$ and $F_{FF}^{res}$ as well as the product $Q_e^{lum} F_{em} = J_0^{SQ}/J_0^{real}$ are easily determined whereas separation of $Q_e^{lum}$ and $F_{em}$ requires the determination of $J_0^{QE}$ by, e.g., an electroluminescence measurement, which is often unavailable for the top-performing cells. Also tabulated are the values $V_{OC}^{real}/V_{OC}^{SQ}$ and $FF_0(V_{OC}^{real})/FF_0(V_{OC}^{SQ})$ which result from the FoMs and are used in Eq. (5) (Box) to calculate the efficiency ratio $\eta_{real}/\eta_0$. Note that for the calculations of the SQ reference values, we used an AM1.5G spectrum and we assumed $n_{id} = 1$ for the calculation of $FF_0$.

| Absorber / Band gap / $E_g^{PV}$ | $\eta_{real}$ / $\eta_{SQ}$ / $\eta_{real}/\eta_{SQ}$ | $J_{SC}^{QE}$ (mA/cm²) / $J_{SC}^{SQ}$ (mA/cm²) / $F_{SC}$ | $FF_0(V_{OC}^{real})$ / $FF_0(V_{OC}^{SQ})$ / $FF_0(V_{OC}^{real})/FF_0(V_{OC}^{SQ})$ | $FF_{cell}$ / $FF_0(V_{OC}^{real})$ / $F_{FF}^{res}$ | $V_{OC}^{real}$ (mV) / $V_{OC}^{SQ}$ (mV) / $V_{OC}^{real}/V_{OC}^{SQ}$ | $F_{em} Q_e^{lum}$ |
|---|---|---|---|---|---|---|
| **GaAs** 1.42 eV | 29.1 % / 33.2 % / **87.7 %** | 29.8 / 32.1 / **92.8 %** | 89.3 % / 89.5 % / **99.8 %** | 86.7 % / 89.3 % / **97.1 %** | 1130 / 1157 / **97.7 %** | **38.7 %** |
| **c-Si** 1.10 eV | 26.7 % / 33.0 % / **80.9 %** | 42.6 / 44.3 / **96.2 %** | 85.2 % / 86.8 % / **98.2 %** | 84.9 % / 85.2 % / **99.7 %** | 740 / 858 / **86.3 %** | **1.1 %** |
| **Cu(In,Ga)Se₂** 1.12 eV | 22.9 % / 33.4 % / **68.6 %** | 38.8 / 43.9 / **88.4 %** | 85.2 % / 87.0 % / **98.0 %** | 79.5 % / 85.2 % / **93.3 %** | 740 / 877 / 84.4 % | **0.57 %** |
| **ABX₃** 1.55 eV | 20.9 % / 31.5 % / **66.4 %** | 24.9 / 27.3 / **91.2 %** | 89.2 % / 90.3 % / **98.8 %** | 74.5 % / 89.2 % / **83.5 %** | 1120 / 1278 / **87.6 %** | **0.24 %** |
| **CdTe₁₋ₓSeₓ** 1.42 eV | 21.0 % / 33.2 % / **63.3 %** | 30.2 / 32.1 / **94.1 %** | 87.0 % / 89.5 % / **97.3 %** | 79.4 % / 87.0 % / **91.3 %** | 880 / 1157 / **76.1 %** | **2.4 × 10⁻³ %** |
| **QD** 1.77 eV | 13.4 % / 27.8 % / **48.2 %** | 15.2 / 20.5 / **74.2 %** | 89.5 % / 91.3 % / **98.0 %** | 76.6 % / 89.5 % / **85.6 %** | 1160 / 1484 / **78.2 %** | **4.9 × 10⁻⁴ %** |
| **OPV** 1.62 eV | 11.2 % / 30.3 % / **37.0 %** | 19.3 / 24.9 / **77.5 %** | 85.8 % / 90.6 % / **94.7 %** | 74.2 % / 85.8 % / **86.5 %** | 780 / 1343 / **58.1 %** | **4.5 × 10⁻⁸ %** |

**Box: Key equations in the SQ model:**

| Equation | Explanation |
|---|---|
| $J = J_{SC} - J_0 \left[ \exp\left(\frac{qV}{n_{id}kT_{cell}}\right) - 1 \right]$ (1) | The current-voltage, $JV$, characteristics of a solar cell are given by a diode current that is increased by the short circuit current $J_{SC}$. Upon relaxing assumptions 1-4, $J_0$, $J_{SC}$, and $T_{cell}$ change from their SQ-values to the real cell values. Here $kT_{cell}$ is thermal energy at the temperature of the solar cell. Note that the ideality factor $n_{id} = 1$ in the SQ-model but may deviate from one in real cells. If solar concentration or emission angle restrictions[29] are considered, it can be taken into account by considering the appropriate values for $J^\circ$ and $J_{SC}$. |
| $V_{OC} = \frac{n_{id}kT_{cell}}{q} \ln\left(\frac{J_{SC}}{J_0} + 1\right)$ (2) | The open-circuit voltage results from solving eqn (1) for $J = 0$. With successive changes from the SQ-case to a real situation, the gain term $J_{SC}$ decreases and the loss term $J_0$ increases, such that $V_{OC}$ decreases from the SQ- to its real solar cell-value. Here, $q$ is the elementary charge. |
| $V_{OC}^{real} - V_{OC}^{SQ} = \frac{kT_{cell}}{q} \ln(F_{SC} F_{em} Q_e^{lum})$ (3) | The loss of open-circuit voltage between that of the real device and the SQ-model, is described by the product of three FoMs (cf. Table 1) $F_{SC} \cdot F_{em} \cdot Q_e^{lum}$ with values < 1. $T_{cell}$ is taken to be that in the SQ-model. |
| $\eta = \frac{P_{max}}{P_{Sun}} = \frac{J_{SC} V_{OC} FF(V_{OC},\ldots)}{P_{Sun}}$ (4) | The efficiency of the solar cell is the maximum output power $P_{max} = (JV)_{max}$, related to the power input $P_{sun}$. Usually, $P_{max}$ is factorized by the product $J_{SC} V_{OC} FF$, where the fill factor $FF$ depends on $V_{OC}$ and various parameters that change the general shape of the diode equation like resistive losses, and $n_{id}$. |
| $\frac{\eta_{real}}{\eta_{SQ}} = F_{SC} \frac{V_{OC}^{real} FF_0(V_{OC}^{real})}{V_{OC}^{SQ} FF_0(V_{OC}^{SQ})} F_{FF}^{res}$ (5) | The ratio between the real efficiency and the SQ-value is defined with the help of four FoMs. Note that $F_{SC}$ enters linearly as well as logarithmically via the open circuit voltage $V_{OC}^{real}$ (cf. Eq. (3)). The equation also considers the change of $FF_0$ from the SQ-case to the real situation as well as the change from $FF_0(V_{OC}^{real})$ to $FF^{real}$, induced by resistive losses with the FoM $F_{FF}^{res} = FF^{real} / FF_0(V_{OC}^{real})$. |
| $FF_0 = \frac{\frac{qV_{OC}}{n_{id}kT_{cell}} - \ln\left(\frac{qV_{OC}}{n_{id}kT_{cell}} + 0.72\right)}{\frac{qV_{OC}}{n_{id}kT_{cell}} + 1}$ (6) | There is no analytical solution for the fill factor $FF$ even in the simplest situation of an ideal diode. However, very precise approximate equations[16] provide a valid relation between $FF_0$, the value without resistive losses, and $V_{OC}$. |

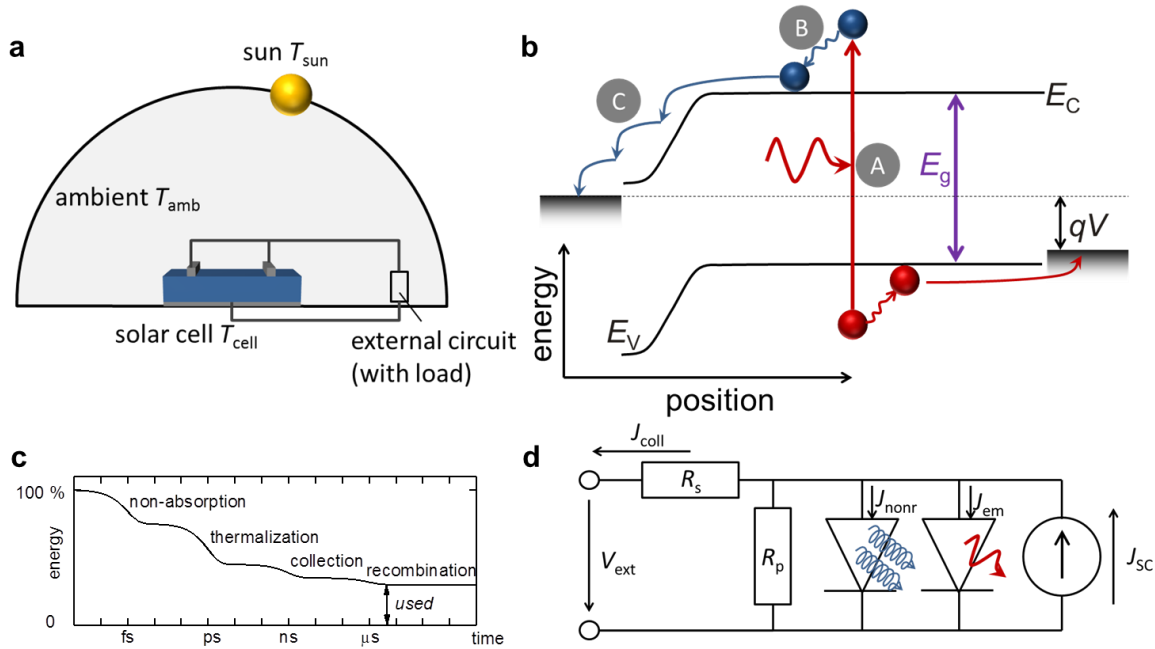

**Fig. 1: a** Schematic of the *world* of the SQ-model, with the sun at $T_{sun}$ illuminating the solar cell at $T_{cell}$ = 300 K, also the temperature of the ambient ($T_{amb} = T_{cell}$), and the external circuit of the solar cell. **b** The three essential steps of photovoltaic power conversion illustrated within an energy band diagram of a *p/n*-junction solar cell; p(n) semiconductor, on right(left). The conversion process consists of three steps, (A) light absorption, (B) local thermalization, directly after photogeneration, and (C) charge collection with further thermalization within the semiconductor and contacts. Selectivity is illustrated as the flow of each of the two types of charge carriers (electrons, blue, and holes, red) into a different contact. **c** Typical time scales of these energy losses, , for illustrative purposes (NB, in Si solar cells, collection and recombination times could be much longer). Non absorption loss, refers to photon whose energy is <$E_g$ . **d** Simple equivalent (electrical engineering) circuit of a solar cell. The SQ-model only requires the current source and the diode (the red photon represents radiative recombination with the current $J_{em}$). Real solar cells are typically described by the addition a second diode, representing non-radiative recombination with current $J_{nonr}$ (indicated by the blue springs, representing heat dissipation), a parallel or shunt resistance $R_p$ and a series resistance $R_s$.

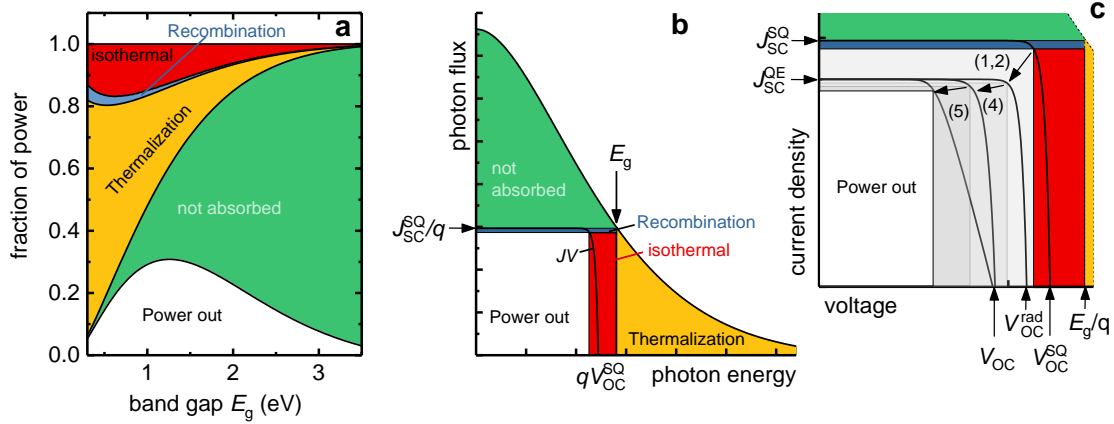

**Fig. 2: a** Illustration of within the SQ-model as function of band gap (always at maximum power point) using for the solar spectrum a 5800 K black body spectrum normalized to 100mW/cm². **b** Energy losses for a given band gap energy depicted as a function of photon flux vs. photon energy (adapted from ref. [10]). By dividing the photon flux by the elementary chare $q$ and multiplying the energy with $q$, the axes can be also read as current density vs. voltage. The black curve (*JV*) denotes the current vs. voltage curve, and the maximum output power is obtained for a maximum area of the white rectangle (Power out), likewise a minimum area for the recombination and isothermal dissipation losses. **c** Current-voltage curves of a solar cell and the power losses occurring upon relaxing the SQ-assumptions 1+2 (combined), 4, and 5. The maximum output power (illustrated by the rectangles) reduces stepwise from the SQ-value $P_{max}^{SQ}$ to the real value $P_{max}^{real}$. The cell temperature $T_{cell}$ is kept at the SQ-value $T_{cell}^{SQ}$ such that assumption 3 is still valid. Violations of assumptions 1-5 are kept at a level such that the real device is still a useful solar cell.

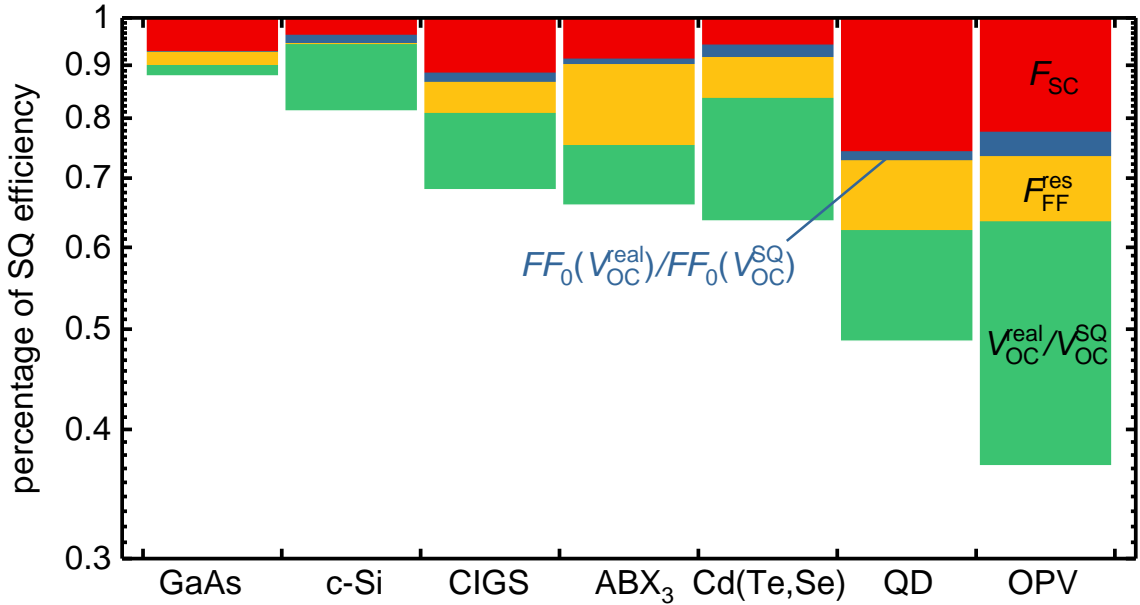

**Fig. 3:** Visualization of potential improvement of top-performing solar cells, based on their band gaps, relative to the ideal, SQ-case, using partitioning of the efficiency losses, according to Eq. (5) (Box). The following figures of merit (FoMs) are used and shown: $F_{SC}$ and $F_{FF}^{res}$, $V_{OC}^{real}/V_{OC}^{SQ}$ and $FF_0(V_{OC}^{real})/FF_0(V_{OC}^{SQ})$. Only the two mature technologies GaAs and Si achieve $\eta_{real}/\eta_0 > 80\ \%$. The poly-crystalline thin-film technologies Cu(In,Ga)Se$_2$ (CIGS), metal-halide-perovskite (ABX$_3$), and Cd(Te,Se) are still below 70 %. Quantum-dot (QD) and organic solar cells (OPV) are below 50 %. Note that for the calculation of $FF_0$, $n_{id} = 1$ was assumed, because $n_{id}$ for record cells is generally not known.